# ARTICLE

# Homogeneous hierarchical NiMoO$_4$@NiMoO$_4$ nanostructure as a high-performance anode material for electrochemical energy storage

Jia Yi Dong, [a] Jin Cheng Xu, [a] Kwun Nam Hui, [a,c] Ye Yang, [a] Xi Tian Zhang, [b] Kar Wei Ng,* [a,c] Shuang Peng Wang, *[a,c] and Zi Kang Tang, * [a,c]

Here we report the extraordinary electrochemical energy storage capability of NiMoO$_4$@NiMoO$_4$ homogeneous hierarchical nanosheet-on-nanowire-arrays (SOWAs) synthesized on nickel substrate by a two-stage hydrothermal process. Comparatively speaking, the SOWAs electrode displays improved electrochemical performances than the bare NiMoO$_4$ nanowire arrays. Such improvements can be ascribed to the characteristic homogeneous hierarchical structure which not only effectively increases the active surface areas for fast charge transfer, but also reduces the electrode resistance significantly by eliminating the potential barrier at the nanowire/nanosheet junction, which is usually an issue in other reported heterogeneous architectures. We further evaluate the performances of the SOWAs by constructing an asymmetric hybrid supercapacitor (ASC) with the SOWAs and activated carbon (AC). The optimized ASC shows excellent electrochemical performances with 47.2 Wh/kg in energy density at 1.38 kW/kg at 0-1.2 V. Moreover, the specific capacity retention can be as high as 91.4% after 4000 cycles, illustrating the remarkable cycling stability of the NiMoO$_4$@NiMoO$_4$//AC ASC device. Our results show that this unique NiMoO$_4$@NiMoO$_4$ SOWAs display great prospect for future energy storage applications.

## Introduction

Supercapacitors (SCs) have shown great promise as the next-generation power devices because of their extraordinary properties, including high-power density, rapid charging and prolonged cycling life. Electrochemical double layer capacitors and pseudocapacitors are the two major types of SCs with different storage mechanisms.[1,2] The electrochemical double layer capacitors stores energy mainly via ion adsorptions at the interface between the electrodes and electrolyte. In comparison, pseudocapacitors have shown advantages in achieving higher energy density and specific capacitance owing to the quick and reversible redox reactions at the surface of the electrodes.[3] The most widely known PCs are fabricated with MnO$_2$ and RuO$_2$. However, the high resistance and dissolution of MnO$_2$ [4,5] in aqueous electrolytes as well as the high cost and toxicity of RuO$_2$ [6,7] hinder them from being used in large scale. In this context, hybrid devices consisting of faradaic and capacitive electrodes arise as a potential candidate which can leverage on the advantages of both types of supercapacitors. To fully exploit the strengths of the hybrid devices, the choice of materials and device structures have to be carefully optimized.

In terms of materials choice, binary metal oxides such as MnMoO$_4$, NiMoO$_4$, NiCo$_2$O$_4$, and CoMoO$_4$ have been extensively investigated owing to their multiple oxidation states as well as significantly better electrical conductivity compared with single-component oxides, rendering them ideal for high-performance charge storage devices.[8,9] Among these binary metal oxides, NiMoO$_4$ has attracted intense research interest due to the superior electrochemical activity of Ni.[3,10] Although Mo does not contribute to the faradic redox reactions which play the most significant role to the overall capacitance/capacity of the device, its high electrical conductivity facilitates the efficient transfer of charges during the process. SCs with respectable specific capacity have been demonstrated with NiMoO$_4$ based nanosheets[11] and nanobundles.[12] To further enhance the electrochemical performances of NiMoO$_4$, a more efficient device architecture is highly desired.

Among various device structures, well-aligned nanoarrays of electrochemically active materials like hydroxides and metal oxides grown on conductive substrates have shown superior advantages in energy storage, especially for SCs.[13,14] Specifically, the nanoarray architecture can provide a large specific surface area for ion diffusion and electron transport, thus resulting in high specific capacity.[15-19] Additionally, growing the active materials directly on the conductive substrates can ensure that the former can adhere to the substrate firmly. This makes the active materials more mechanically stable during the charge/discharge process.[20,21] 3-dimensional (3D) structures have been introduced to further enhance the energy storing capability of nanoarrays by increasing the active area for redox

a. *Joint Key Laboratory of the Ministry of Education, Institute of Applied Physics and Materials Engineering, University of Macau, Avenida da Universidade, Taipa, 999078, Macao SAR, China.*

b. *Department of physics, Harbin Normal University, 150000, Harbin, China.*

c. *Department of Physics and Chemistry, Faculty of Science and Technology, University of Macau, Taipa, 999078，Macao SAR, China.*

reactions.[22-25] Heterogeneous SOWAs, a type of hierarchical nanoarrays constructed with different materials in the nanowire core and nanosheet shell, have been widely studied for realizing high-performing SCs.[26-28]

Compared with the heterogeneous architecture, homogeneous SOWAs show great promise in lowering the intrinsic resistance by eliminating the surface potential barrier at the core/shell junction.[29, 30] In spite of the potential advantages, very few works have been reported on the rational design of homogeneous SOWAs for SCs. Therefore, the pioneering study of such architecture on $NiMoO_4$ is highly desirable. In this paper, we report the fabrication and characterizations of a novel homogeneous $NiMoO_4@NiMoO_4$ hierarchical SOWAs supercapacitors. The SOWAs, comprising $NiMoO_4$ nanowires covered with $NiMoO_4$ nanosheets, are realized d on a Ni substrate by a two-stage hydrothermal treatment. Impressively, the SOWAs show over 30% enhancements in specific capacity and noticeably better electrochemical behaviors comparing with the electrodes formed by bare $NiMoO_4$ nanowire arrays. We do believe that the significance reduce in series resistance of our homojunction material is benefit for charge carrier transport along the interconnected nanowire network. Observing these extraordinary properties, we further utilized the homogeneous SOWAs structure to fabricate ASC which display excellent cycling stability and electrochemical performances superior to other reported devices with heterogeneous hierarchical structures. These promising results fully demonstrate the potential for the $NiMoO_4@NiMoO_4$ SOWAs to be used in energy storage applications which require low resistive loss, fast operations and good mechanical stability.

## Results and discussion

The crystallographic phase of the three samples is first assessed using XRD (Fig. 1a). Owing to the strong signal coming from the Ni substrate, the diffractions of the SOWAs cannot be readily resolved in Fig. 1a. When zooming in to the regime between 20° and 40° (Fig. 1b), one can clearly identify three diffractions which can be assigned to monoclinic $NiMoO_4$. These diffractions are consistent with JCPDS data (Card No.86-0361),[31] exemplifying the purity of the $NiMoO_4$ synthesized with the hydrothermal processes in all samples.

It is well known that rod-like morphology can provide more reaction sites and improve ion transport.[32-34] To verify that our samples are indeed in the desired architecture, we extensively

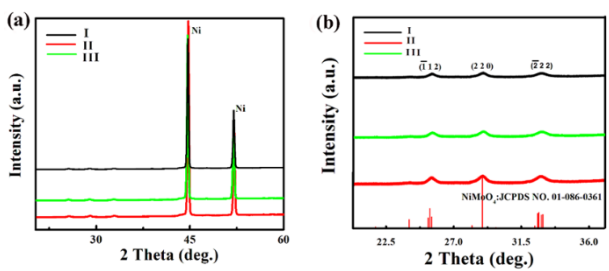

**Fig. 1** (a) XRD of Sample I, Sample II and Sample III grown on Ni foam; (b) Zoom-in view of the XRD peaks from 20°~ 40°.

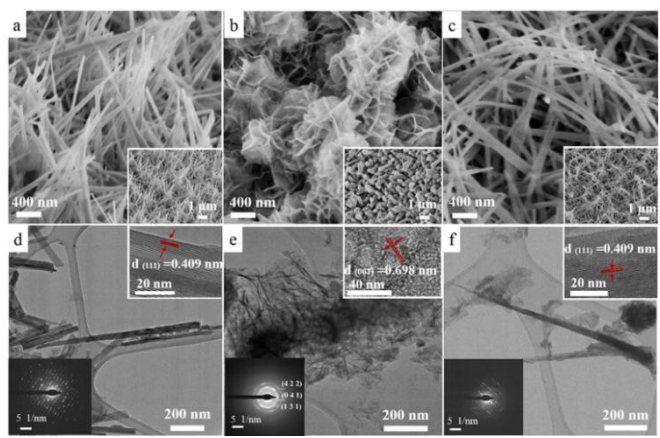

**Fig. 2** FESEM images (a-c) of Samples I, II and III. The insets show a large-area view of each sample. TEM images (d-f) of three samples. The insets in each TEM image show the corresponding SAED pattern and HRTEM image.

studied their morphology with SEM and TEM. Figs. 2a-c show representative SEM images revealing the morphologies of the three samples. From Fig. 2a, we can see that $NiMoO_4$ in Sample I grow into rods which generally align to the vertical direction and form an open network which covers the entire substrate surface uniformly. Such nanowire framework provides a large area for redox reactions and charge storage. These capabilities are further enhanced by introducing the extra secondary $NiMoO_4$ nano-flakes in sample II. Impressively, as shown in Fig. 2b, the $NiMoO_4$ nanowire surface is completely covered with $NiMoO_4$ nanosheet layers. These nanosheets significantly increase the density of active sites, thus enabling the full utilization of the active materials for energy storage. In Sample III, the extra mass of $NiMoO_4$ for forming the 3D nanocomposite shell in Sample II is incorporated into the growth of nanowire arrays. As illustrated in Fig. 2c, the nanowires become too long and lose the desired orthostatic morphology. The results here confirm that Sample II, an orthostatic nanowire network decorated with ultra-thin nanosheets, should present the ideal morphology for SC applications.

To get a clearer idea of the nanoscopic crystallinity of the hierachical structures, we performed extensive TEM analysis. Fig. 2d shows a typical 100-nm-thick $NiMoO_4$ nanowire obtained from Sample I. The high-resolution TEM (HRTEM) image in the inset reveals clear lattice fringes exhibiting a periodicity of around 0.41 nm, which is in excellent agreement with the interplanar distance of the (111) planes of $NiMoO_4$. The single-crystalline nature is further demonstrated with the selective area electron diffraction (SAED) pattern in Fig. 2d. The excellent crystallinity is crucial to efficient charge exchange and transport along the wires. Fig. 2e demonstrates the architecture of Sample II comprising of overlapping thin nanosheets, which agrees well with the SEM image in Fig. 2b. Moreover, the SAED in Fig. 2e exemplifies the polycrystallinity of Sample II with clearly indexable 4 2 2, 0 4 1 and 1 3 1 diffraction rings. The HRTEM image further confirms the short-term crystallinity of the nanosheets which display lattice fringes separated by around 0.698 nm, which matches the (001) interplanar spacing NiMoO$_4$. The lattice in Sample III (Fig. 2f) are similar to those in Sample I, except that nanowires of Sample III are longer. Elemental mapping by energy dispersive spectroscopy (Fig. S1) of the central region of Sample II demonstrates that the nanostructures are chemically composed of Mo, O and Ni which are uniformly distributed over the entire sample. The TEM studies elucidate the chemical composition uniformity and crystalline properties of the NiMoO$_4$ SOWAs, which are crucial to the electrochemical performances of the novel material system.

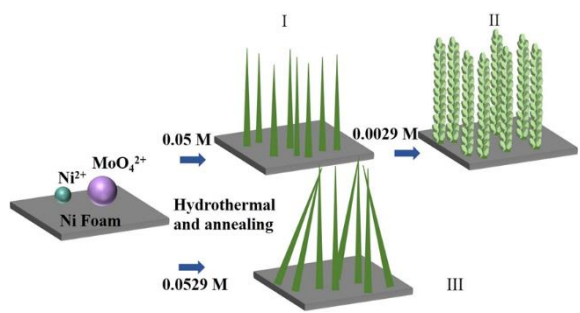

**Fig. 3** Illustration of the growth mechanism of Samples I, II and III.

The electron microscopy investigations imply that the SOWAs should possess the largest surface areas for charge exchange. We verify this by measuring the surface areas of the three samples quantitatively by BET measurement and the results are plotted in Fig. S2. A BET surface area of 33.2 m$^2$/g was observed for Sample II, which is over 20% larger than the surface areas of Sample I (26.2 m$^2$/g) and Sample III (27.7 m$^2$/g), as revealed shown in the N$_2$ adsorption-desorption profiles. The narrow hysteresis loops that appeared at different relative pressures are characteristics of hierarchical pores.[35] The relatively high specific surface area in sample II not only increases the electrode/electrolyte contact area but also provides a lot more effective sites for redox reactions. These unique properties indicate that the SOWAs should exhibit superior electrochemical performance, which is to be discussed in more details below.

Based on the electron microscopy studies, we developed a growth mechanism as illustrated in Fig. 3. First of all, a calcination process after a hydrothermal reaction gives rise to a densely packed nanowire array in Sample I. Subsequently, interconnected NiMoO$_4$ nanosheets with various lateral sizes are grown on the NiMoO$_4$ nanowire framework via a secondary hydrothermal treatment, resulting in the hierarchical structure in Sample II after a secondary calcination process. If all the reactants were supplied at the same time, the resulting nanowires would lose the desirable orthostatic property as in Sample III.

Hydrothermal reaction time and the concentration of reactants undoubtedly play an important role in the morphology of the metal oxide electrodes, and thus affect their electrochemical performance. To validate the advantages of NiMoO$_4$@NiMoO$_4$ SOWAs, we investigated the electrochemical performances of the three samples. Electrodes I, II and III were fabricated with Samples I, II and III, respectively. Fig. 4a displays the representative cyclic voltammetry (CV) profiles of the three electrodes within the voltage range of 0-0.8 V (vs. Hg/HgO) at 30 mV/s. Notably, Electrodes I and III show similar CV behaviors, indicating that the slightly larger mass in Electrode III does not have big impact on the capacitance. In contrast, Electrode II shows more distinct redox peaks in its CV curve. Besides, the obviously larger enclosed CV curve area in Electrode II than those in Electrodes I and III at identical sweep speed implies that the former has much respectably higher specific capacitance. Indeed, the discharging time is ~ 60% longer in Electrode II compared with the other two at 1 A/g in the GCD plot in Fig. 4b. Although the CVs in Fig. 4a were scanned within a voltage range of 0 to 0.8V, we set the voltage range for GCD at 0-0.52 V because considerable polarization is observed at potential higher than 0.52V.[3, 36] To avoid such undesirable effect, it is an usual practice to offset the voltage range of GCD from that used in the CV profiles.[36-39]

The ultra-long discharge time elucidates the excellent electrochemical performance of Electrode II in accordance with equations (1) and (2). To further investigate the electrochemical properties of Electrode II, we measured its CV at various sweep rates and GCD at various current levels. As represented in Fig. 4c, all the CV profiles reveal meristic and clear redox peaks, which are strong indications that the specific capacitance characteristics predominantly originate from the faradic redox reactions of Ni$^{2+}$/Ni$^{3+}$ (Fig. 4a, c).[40, 41] In addition, the redox peaks are prominent under low scanning rates (5 and 10 mV/s) but weaken at high sweep rates due to the slow diffusion of OH$^{-1}$ at the interface of electrolyte/electrode.[42] Hence, C$_{SC}$ and C$_{AC}$ obtained at the lowest scan rate can be considered as the closest to complete utilization of the electrode. Nevertheless, no significant change in the shape of the CV curves is observed

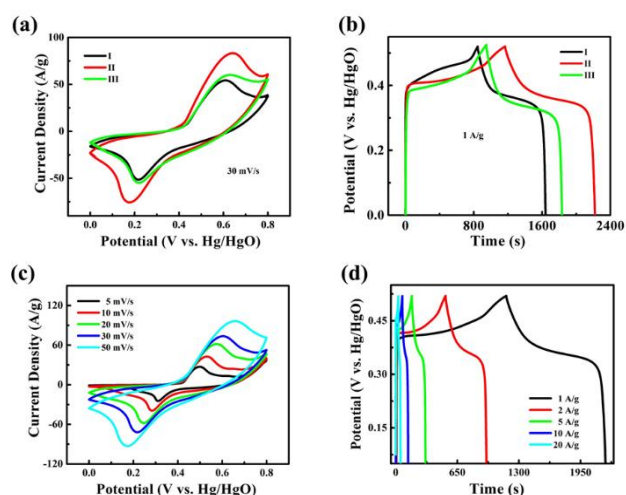

**Fig. 4** (a) CV profiles of Electrodes I, II and III at 30 mV/s; (b) GCD profiles of Electrodes I, II and III at 1 A/g; (c) CV profiles of Electrode II at different sweep speeds; (d) GCD profiles of Electrode II at various current.

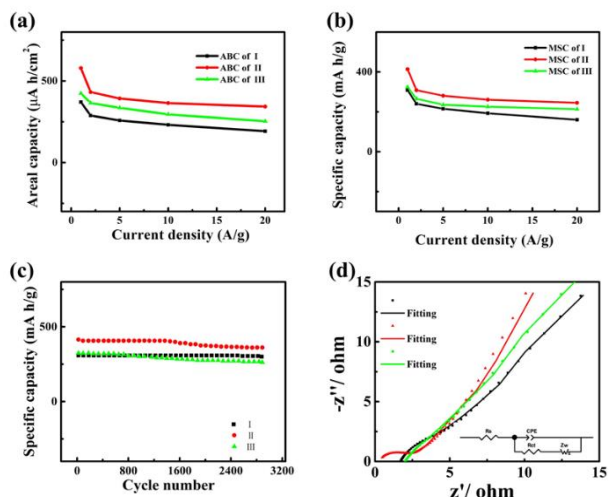

**Fig. 5** Electrochemical performances of the three Electrodes I, II and III. (a), (b) Areal and specific capacity behavior at various current density; (c) Cycling behavior at 1 A/g; (d) The Nyquist plots and fitting curves in the frequency range of 0.01-100 kHz and the equivalent circuit diagram used for the analysis of EIS data.

when the scan rate increases and the peak current still increases proportionally, indicating that Electrode II can facilitate fast redox reactions. Additionally, the highly symmetric GCD profiles obtained at different current densities varying from 1 to 20 A/g (Fig. 4d) clearly elucidates the excellent electrochemical behavior and reversible redox reaction activity.

Based on the above GCD behaviors, we calculate the areal capacity and specific capacity of the three electrodes and plot out their dependences on the current densities, as shown in Fig. 5 (a-b). Since the effective contact area between the active materials and the electrolyte significantly diminishes at high sweep rate, the capacity drops upon increasing current density.[43-45] The capacity of Electrode II can reach up to 413 mA h/g (578 μA h/cm$^2$) at 1 A/g and drops to around 46.7% of the maximum value, i.e. 220 mA h/g (308 μA h/cm$^2$) at 20 A/g. The maximum capacities of Electrodes I and III are much lower and can only reach 309 mA h/g (371 μA h/cm$^2$) and 324 mA h/g (453 μA h/cm$^2$), respectively, which are over 25% and 21% lower than that of Electrode II. Notably, the specific capacity of our homogeneous architecture here is favorably comparable with those of previously reported nanostructures, including CoMoO$_4$@Co(OH)$_2$ core-shell structures (265 mA h/cm$^2$ at 2 mA/cm$^2$),[46] as well as CoMoO$_4$ nanoflakes (32.40 mA h/g; 492.48μAh/cm$^2$),[47] Ni-Mo-S nanosheets (312 mA h/g at 1 mA/cm$^2$),[48] and flower-like Mn-Co oxysulfide (136 mA h/g at 2 A/g).[49] Detailed comparisons in electrochemical performances among these devices and our structure can be found in Table S1. The superior energy storage performance of the NiMoO$_4$@NiMoO$_4$ SOWAs might be due to the larger specific surface area as well as more efficient charge carrier transport across the homojunction in our homogeneous hierarchical architecture.

The cycling stability of the three electrodes was assessed by GCD with the voltage ranging from 0-0.52 V, as displayed in Fig. 5c. The capacity of electrodes I, II and III respectively drop by 1.5%, 4.3% and 6.9% to 300.3 mA h/g, 361.2 mA h/g and 262.3 mA h/g after 3000 cycles at a discharge current density of 1 A/g. Surprisingly, Electrode II experienced a relatively large cycling degradation, which could be attributed to the lack of stability in the structure of the NiMoO$_4$ nanosheets upon high current operations. Nevertheless, the specific capacity of Electrode II stabilizes after around 3000 cycles and this value is still respectably higher than those of Electrode I and III.

Aside from surface area, the impedance of the electrode also has an important effect on the overall behavior of SCs. The impedance properties of the electrodes were investigated in an open-circuit voltage with 5 mV amplitude utilizing an EIS measurement.[15, 29, 50, 51] Fig. 5d, displays the Nyquist plots measured in the frequency range of 0.01-100 kHz and the equivalent circuit is shown in the inset. $R_s$, CPE, $R_{ct}$ and $Z_w$ correspond to the equivalent series resistance (ESR), constant phase element, charge-transfer resistance and Warburg impedance, respectively. These elements basically define the various resistances arose in the redox reactions. From the Nyquist plot, one can obtain the ESR from the X-intercept at the high frequency regime, which is a combined effect from the electrolyte resistance, the active material resistance and the contact resistance between the electrode and electrolyte.[50] Apparently, Electrode II displays the smallest ESR which is only 0.16 Ω. This very small ESR is more than 10 times lower than those of Electrode I (1.74 Ω) and Electrode III (2.03 Ω). It was also observed that the ESR of our homogeneous structure decreased at least 68% compared with other heterogeneous structures or different morphologies of NiMoO$_4$. (Table S1) Moreover, the diameter of the semicircle represents the interfacial charge-transfer resistance ($R_{ct}$).[51] In our case, the $R_{ct}$ of Electrode I, II and III are 3.0 Ω, 2.35 Ω and 2.65 Ω, respectively, exemplifying that Electrode II has the best charge-transfer characteristics. Finally, the Nyquist plot of Electrode II exhibits a slope drastically steeper than those of Electrode I and III at low frequencies. This validates the much better capacitive performance of Electrode II as a result of respectably lower diffusion resistance. These results suggest that the significantly better capability and stability of our homogeneous hierarchical structure can mainly be ascribed to the low resistances.

To assess the feasibility of the of the superior electrochemical properties of NiMoO$_4$@NiMoO$_4$ SOWAs for practical applications, we fabricated an asymmetric hybrid supercapacitor using the AC and optimized SOWAs as the negative and positive electrodes, respectively. The CV profiles of the ASC within 0-1.2 V at various sweep rates (5 to 50 mV/s)

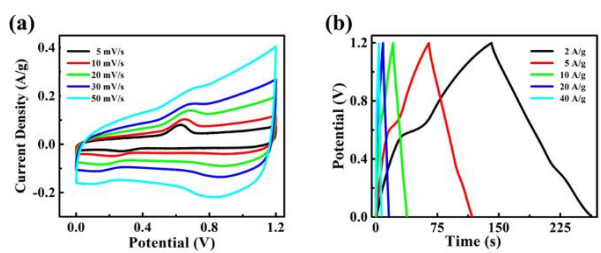

**Fig. 6** Electrochemical performance of NiMoO$_4$@NiMoO$_4$ SOWAs//AC ASC. (a) CV curves at various scan rates; (b) GCD curves at various current densities.

**4** |

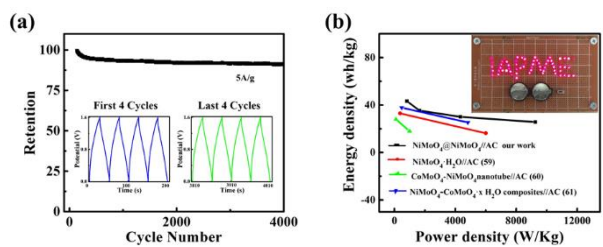

**Fig. 7** NiMoO$_4$@NiMoO$_4$//AC (a) 4000 cycles cycling performance with the inset GCD plots of the first and last four circles (b) The comparison of the Ragone plot with previously ASCs, and the inset picture shows two NiMoO$_4$@NiMoO$_4$//AC devices connected in series lighting up 64 red LEDs.

were displayed in Fig. 6a. And the CV profiles were tested at a constant scan rate of 30 mV/s within various range of voltage (see Fig. S3). The peak tail, which appears as a consequence of undesirable oxygen evolution reaction-induced peak in the CV profiles when the voltage is broaden to 1.5 V or higher, indicates that the maximum working voltage of the ASC should be around 1.2 V. Such phenomenon was explained in similar way in the previous literatures. [52-54] For example, Xu et al. reported that the optimized operating voltage range of the ZIF-LDH/GO//ZIF-C/G device was 0-1.6 V in order to avoid the oxygen evolution reaction at 1.6-1.8 V.[55] Elshahawy et al. reported that 1.6 V was utilized to be the upper limit of the operating voltage since when the voltage went above 1.7 V, the oxygen evolution of Co(P,S)/CC//PCP/rGO device started.[56] It can be clearly seen that the overall capacitances of the NiMoO$_4$@NiMoO$_4$//AC ASC is composed of two components, namely the EDLC-type capacitance and Faradaic pseudocapacitance.[50] The shape of CV profiles shows no distortion as the scan rate increases, indicating the desirable charging and discharging behaviors. The performance of GCD (Fig. 6b) was also evaluated while the voltage of ASC reached 1.2 V at a current density in between 2 and 40 A/g. The areal and specific capacity of the ASC are estimated on the basis of the discharge profiles, which are substantially higher than those ASCs reported before, for instance, ZnCo$_2$O$_4$@MnO$_2$//α-Fe$_2$O$_3$ (0.40 F/cm$^2$ at 2.5 mA/cm$^2$), Co$_{0.85}$Se//AC (0.33 F/cm$^2$ at 1 mA/cm$^2$), and NiCo$_2$O$_4$@Co$_{0.33}$Ni$_{0.67}$(OH)$_2$//CMK-3 (0.89 F/cm$^2$ at 5 mA/cm$^2$).[57,58]

Cycling Characteristics is another crucial factor that influences the performance of SCs. Fig. 7a elucidates that the ASC device shows outstanding cycling stability. In particular, a 91.4% retention rate of the initial capacitance is obtained after 4000 cycles at a current density of 5 A/g. Furthermore, the repeated GCD profiles of the last 4 cycles show identical shape as the profiles of the first 4 cycles, as displayed in the inset of Fig. 7a). The excellent capacitance stability of the ASC reveals that the NiMoO$_4$@NiMoO$_4$ SOWAs is suitable as a stable electrode material.

To obtain a more comprehensive assessment of electrochemical SCs, energy and power densities are also two important components that should be taken into account. Fig. 7b shows the Ragone plot which correlates the energy density with the power density of ASC devices. The energy density of Sample II, i.e. NiMoO$_4$//AC ASC, can achieve 47.2 Wh/kg at 1.38 KW/kg, and still retains 25.7 Wh/kg at 9.25 KW/kg. Additionally, the energy density of our devices is compared with NiMoO$_4$//AC ASC reported previously. As seen in Fig. 7b, our devices show considerably higher energy density than NiMoO$_4$·H$_2$O//AC (17.72 Wh/kg),[59] CoMoO$_4$-NiMoO$_4$ nanotube//AC (33 Wh/kg at 375 W/kg),[60] β-NiMoO$_4$-CoMoO$_4$·xH$_2$O composites//AC (28 Wh/kg at 100 W/kg).[61] The inset image of Fig. 7b shows a demonstration of our ASC used in real operations. Two ASCs connected in series were encapsulated in a battery case and used to light up 64 red light emitting diodes connected in parallel for 137 seconds. The exciting results presented here fully elucidate the extraordinary electrochemical energy storage capability of the NiMoO$_4$@NiMoO$_4$//AC ASC.

## Experimental

### Materials Synthesis.

Rectangular nickel conductive substrates (2.4 cm × 3 cm) were ultrasonically cleaned with HCl acid (6 M), ethanol and deionized (DI) water, each for fifteen minutes sequentially. NiMoO$_4$ nanowire arrays were grown on a Ni substrate using a single step hydrothermal treatment. The synthesis typically starts with adding an aqueous solution of 0.05M Ni(NO$_3$)$_2$·6H$_2$O and 0.05M Na$_2$MoO$_4$ solution under constant magnetic stirring. Nanowires with the desired density and aspect ratio were found to grow when a 25 vt % ethanol solvent was added. Then, the Ni substrate was immersed into the mixture in a 40 mL autoclave and kept at 140 °C for 6 h. Subsequently, the substrate was removed from the solution and washed with DI water. The sample was annealed for 2 h at 400 °C in flowing argon afterwards. The sample obtained with this process is denoted as Sample I.

### Preparation of SOWAs.

To synthesize the aligned SOWAs, the nanowire arrays in sample I was utilized as the backbone for the growth of nanosheets via a secondary hydrothermal reaction. Briefly, an equivalence of Sample I was loaded into an autoclave containing a mixture of 0.0029 M of Ni(NO$_3$)$_2$·6H$_2$O and 0.0029 M of Na$_2$MoO$_4$. The reactants were then kept for 3 h at 140 °C, followed by washing and annealing in conditions similar to those used in Sample I. This SOWAs are denoted as Sample II.

### Preparation of Nanowire Arrays with Same Mass as Sample II.

To illustrate that the difference in electrochemical performance between Samples I and II is not due to the slight mass increase of NiMoO$_4$ in the latter, we fabricated another nanowire array with the same nominal mass as Sample II. The process started with loading a Ni substrate into an solution that consist with Ni(NO$_3$)$_2$·6H$_2$O and Na$_2$MoO$_4$ equally at 0.0529 M, then heated up for 9 h at 140 °C. The resulting nanowire arrays were then washed and annealed in conditions similar to those used in Sample I. The obtained sample was denoted as Sample III.

### Preparation of an AC Electrode.

Commercial activated carbon, acetylene black and conducting graphite with mass fractions of 80 wt %, 7.5 wt % and 7.5 wt %, respectively, were mixed to obtain a homogeneous black powder. 5 wt % of poly(tetrafluoroethylene) and certain amount of ethanol were added subsequently. The final paste was then fixed onto a Ni substrate under 10 MPa and dried for 12 h at 80 °C.

**Materials Characterization.**

The morphology and crystal structure were characterized by field emission scanning electron microscopy (FESEM, Sigma, Zeiss), X-ray diffraction (XRD, Smartlab, Rigaku) and transmission electron microscopy (TEM, Talos F200X, FEI). The surface area was calculated using the Brunauer–Emmett–Teller (BET, 3Flex, Micromeritics) method within a relative pressure (P/P$_0$) range of 0.05-0.45.

**Electrochemical Measurements.**

Electrochemical performances such as the cyclic voltammetry (CV) and galvanostatic charge-discharge profiles (GCD) were obtained with an electrochemical analyzer (CHI 760E, Shanghai Chenhua) at ambient temperature. The measurements were performed in a three-electrode electrochemical cell which contains a 1 M KOH aqueous solution (electrolyte), a standard Hg/HgO (reference electrode) and Pt foil (counter electrode). We carried out the electrochemical impedance spectroscopy (EIS) measurements utilizing an alternating-current (AC) source which can generate a voltage of 10 mV amplitude varying in a frequency ranging between 0.01 kHz and 100 kHz. Several CV cycles were performed as an activation step before the actual data collection. Samples I, II and III were employed directly as the working electrodes. The effective working area of the electrode, i.e. the area of immersion inside the electrolyte, was fixed at 1 cm$^2$. We estimate the mass loading to be approximately 1.2 mg/cm$^2$, 1.4 mg/cm$^2$, 1.4 mg/cm$^2$, respectively. Specific capacity (C$_{SC}$, mA h/g) and areal capacity (C$_{AC}$, mA h/cm$^2$) of the three electrodes were examined in three-electrode system by equations (1) and (2):

$$C_{SC} = \frac{2I \times \int v\, dt}{mV} \quad (1)$$

$$C_{AC} = \frac{2I \times \int v\, dt}{SV} \quad (2)$$

where m and S are the mass and area of the active electrodes, I is the discharge current, and ∫ Vdt represents the area enclosed under the discharge curves.

For the two-electrode ASCs systems, equations (3) and (4) were used to calculate the energy and power densities, respectively. The mass loading of AC was around 2.8 mg/cm$^2$ according to equation (5).

$$E = \frac{I \times \int v\, dt}{m \times 3.6} \quad (3)$$

$$P = \frac{E}{t} \times 3600 \quad (4)$$

$$\frac{m_+}{m_-} = \frac{C_{S-} \Delta V_-}{C_{S+} \Delta V_+} \quad (5)$$

## Conclusions

In summary, uniform 3-dimensional NiMoO$_4$@NiMoO$_4$ SOWAs have been successfully demonstrated using a 2-step hydrothermal and calcination processes. The unique hierarchical architecture exhibits enhanced electrochemical behaviors, which can be ascribed to the increased reaction area and lower series resistance for carrier transport. In addition, the ASC device constructed with the SOWA demonstrates an impressive energy density of 47.2 Wh/kg at a power density of 1.38 kW/kg. These exciting results clearly indicate that the homogeneous SOWAs system can be used practically in constructing SCs with superior performance. In addition, the extraordinary synthesis tactics may have good guidance to the construction of 3D nanostructures for implementing electrodes which hold great promise for practical applications in electrochemcial energy storage.

## Conflicts of interest

The authors declare no competing financial interest

## Acknowledgements

This work was supported by Science and Technology Development Fund from Macau SAR (FDCT084/2016/A2, FDCT051/2017/A and FDCT199/2017/A3) and Multi-Year Research Grants (MYRG2017-00152-FST, MYRG2017-00149-FST, SRG2016-00085-FST, SRG2016-00073-FST, SRG2016-00002-FST) from the Research Services and Knowledge Transfer Office at the University of Macau.

# Supporting Information

# Homogeneous hierarchical NiMoO$_4$@NiMoO$_4$ nanostructure as a high-performance anode material for electrochemical energy storage

Jia Yi Dong,[a] Jin Cheng Xu,[a] Kwun Nam Hui,[a,c] Ye Yang,[a] Xi Tian Zhang,[b] Kar Wei Ng,*[a,c] Shuang Peng Wang,*[a,c] and Zi Kang Tang,*[a,c]

[a] *Joint Key Laboratory of the Ministry of Education, Institute of Applied Physics and Materials Engineering, University of Macau, Avenida da Universidade, Taipa, 999078, Macao SAR, China*

[b] *Department of physics, Harbin Normal University, 150000, Harbin, China*

[c] *Department of Physics and Chemistry, Faculty of Science and Technology, University of Macau, Taipa, 999078, Macao SAR, China*

**\*** Prof. Kar Wei Ng, E-mail: billyng@umc.mo

**\*** Prof. Shuang Peng Wang, E-mail: spwang@umc.mo

**\*** Prof. Zi Kang Tang, E-mail: zktang@umc.mo

# Content

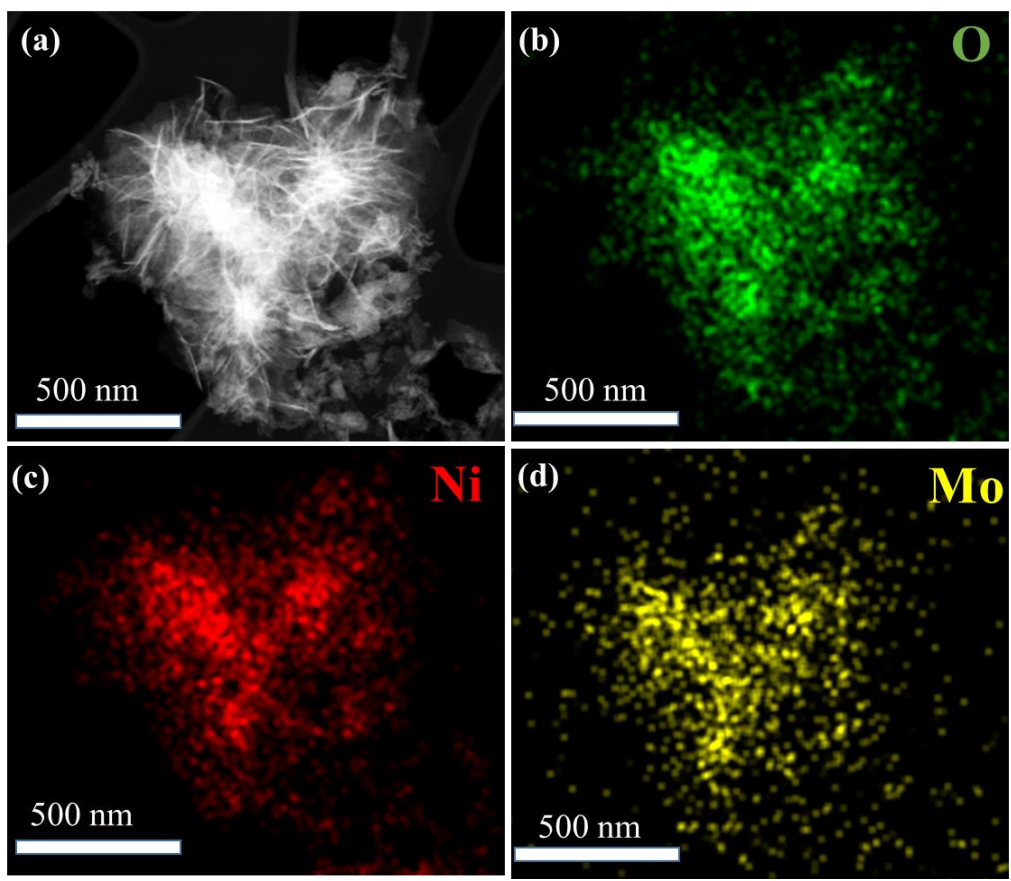

**Figure S1** Images of the Sample II. High angle annular dark field (HAADF) TEM image (a) and (b-d) corresponding elemental mapping, revealing the homogeneous distribution of O, Ni and Mo in the sample.

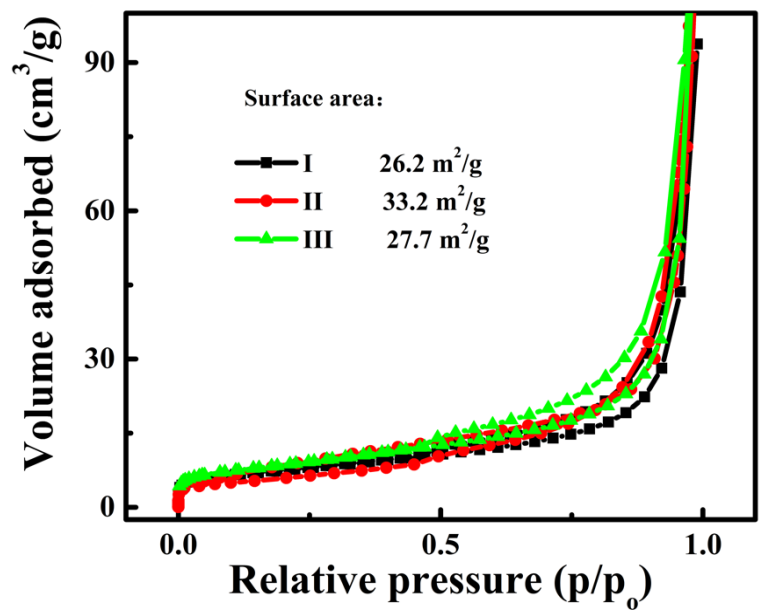

**Figure S2** N₂ absorption-desorption isotherms of Samples I, II, III.

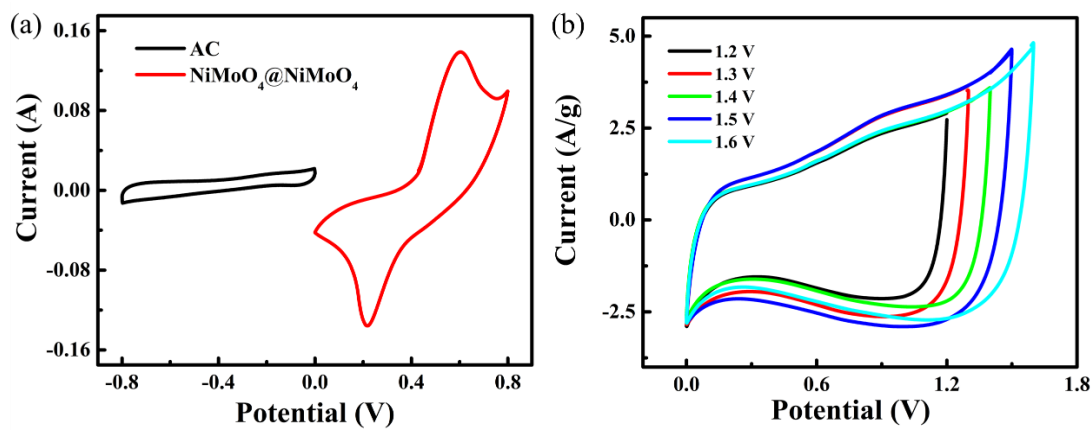

**Figure S3** (a) CV curves of NiMoO₄@ NiMoO₄ SOWAs and AC electrode with a scan rate of 30 mV/s; (b) CV curves of the ASC device collected at different scan voltage windows.

**Table S1** Comparison among our device with other heterostructure electrode materials.

| Electrode materials | Specific/areal capacity/capacitance | Cyclic stability | Energy density | Rs | Ref. |
|---|---|---|---|---|---|
| $CoMoO_4$@Co$(OH)_2$ core-shell | 265 mA h/cm$^2$ at 2 mA/cm$^2$ | 98.64% retention after 2500 cycles | — | — | 1 |
| $CoMoO_4$ Nanoflakes | 32.40 mA h/g (492.48 µAh/cm$^2$) | 85.98% retention after 5000 cycles | — | 1.8 Ω | 2 |
| Ni-Mo-S nanosheets | 312 mA h/g at 1 mA/cm$^2$ | 95.86% retention after 10000 cycles at 50 mA/cm$^2$ | 82.1 W h/kg at 0.56 kW/kg | — | 3 |
| Flower-like Mn-Co oxysulfide | 136 mA h/g at 2 A/g | 86.5% retention after 3000 cycles at 20 A/g | — | — | 4 |
| $NiMoO_4$@Ni$(OH)_2$ core/shell | 7.43 F/cm$^2$ at 4 mA/cm$^2$ | 72% retention after 1000 cycles at 8 mA/cm$^2$ | — | 0.5 Ω | 5 |
| $NiMoO_4$ nanospheres | 974.4 F/g at 1 A/g | 64% retention after 2000 cycles at 5 A/g | 20.1 Wh/kg at 2100 W/kg | 0.61 Ω | 6 |
| β-$NiMoO_4$-$CoMoO_4$·x$H_2O$ | 1472 F/g at a 5 mA/cm$^2$ | 92% retention after 1000 cycles at 10 mA/cm$^2$ | 28 Wh/kg at 100 W/kg | 2.5 Ω | 7 |
| $NiMoO_4$@Co$(OH)_2$ core/shell | 2.335 F/cm$^2$ at 5 mA/cm$^2$ | 83% retention after 5000 cycles at 20 mA/cm$^2$ | — | 2.0 Ω | 8 |
| $MnO_2$@$NiMoO_4$ core-shell | 186.8 F/g at 10 mV/s | 132.7% retention after 20000 cycles at 80 mV/s | 32.5 Wh/kg at 0.75 KW/kg | 1.5 Ω | 9 |
| $MnCo_2O_4$@$MnO_2$ core-shell | 858 F/g at 1 A/g | 91% retention after 5000 cycles 2 A/g | 6.0 Wh/kg at 252 W/kg | 4.8 Ω | 10 |
| $NiMoO_4$@$NiMoO_4$ SOWAs | 413 mA h/g (578 µAh/cm$^2$) at 1 A/g | 91.4 % retention after 4000 cycles at 5 A/g | 47.2 Wh/kg at 1.38 kW/kg | 0.16 Ω | Our work |